\documentclass[aps,prl,reprint,showpacs,notitlepage,superscriptaddress,twocolumn]{revtex4-2}
\usepackage{amssymb}
\usepackage{mathrsfs}
\usepackage{amsfonts}
\usepackage{graphicx}
\usepackage{amsmath}
\usepackage{dcolumn}
\usepackage{color}
\usepackage{subfigure}
\usepackage{epsfig}
\usepackage{soul}
\usepackage[usenames,dvipsnames]{xcolor}
\usepackage[colorlinks,linkcolor=blue,citecolor=blue,hyperindex,bookmarks=false,pdfstartview=FitH]{hyperref}
\usepackage{bm}
\usepackage{dsfont}
\usepackage{verbatim}
\usepackage{multirow}
\usepackage{mathrsfs}
\usepackage{leftidx}
\usepackage{xspace}
\usepackage{braket}
\usepackage{bbm}
\usepackage{cancel}
\usepackage[normalem]{ulem}
\providecommand{\U}[1]{\protect\rule{.1in}{.1in}}

\newcommand{\figpanel}[2]{\hyperref[#1]{\ref*{#1}(#2)}}

\definecolor{alanred}{RGB}{102, 0, 255}

\usepackage{mleftright} 
\usepackage{siunitx}

\begin{document}

\title{Dressed Interference in Giant Superatoms: Entanglement Generation and Transfer}
\author{Lei Du}
\affiliation{Department of Microtechnology and Nanoscience (MC2), Chalmers University of Technology, 412 96 Gothenburg, Sweden}
\author{Xin Wang}
\affiliation{Institute of Theoretical Physics, School of Physics, Xi'an Jiaotong University, Xi'an 710049, China}
\author{Anton Frisk Kockum}
\affiliation{Department of Microtechnology and Nanoscience (MC2), Chalmers University of Technology, 412 96 Gothenburg, Sweden}
\author{Janine Splettstoesser}
\affiliation{Department of Microtechnology and Nanoscience (MC2), Chalmers University of Technology, 412 96 Gothenburg, Sweden}

\begin{abstract}
We introduce the concept of giant superatoms (GSAs), where two or more interacting atoms are nonlocally coupled to a waveguide through one of them, and explore their unconventional quantum dynamics. For braided GSAs, this setup enables decoherence-free transfer and swapping of their internal entangled states. For separate GSAs, engineering coupling phases leads to state-dependent chiral emission, which enables selective, directional quantum information transfer. This mechanism further facilitates remote generation of $W$-class entangled states. Our results thereby open exciting possibilities for quantum networks and quantum information processing.
\end{abstract}

\date{\today}

\maketitle

\emph{Introduction.---}Giant atoms---quantum emitters coupled to their environments at multiple spatially separated points---have reshaped our understanding of light-matter interactions~\cite{fiveyears}. In particular, they challenge the conventional understanding that interactions between single atoms and fields should be treated in a point-like and local manner. Experimentally, giant atoms have been realized using superconducting qubits coupled to surface acoustic waves~\cite{SAW2014,nonexp} or to meandering microwave transmission lines~\cite{OliverBraided,WilsonPRA2021,JoshiPRX2023}. 
In addition, a number of feasible implementations have been proposed, relying on building blocks such as dynamically modulated optical lattices~\cite{AGT2019GA}, coupled-waveguide arrays~\cite{LonghiGA,Lim2023PRA}, Rydberg-atom pairs~\cite{CYT2023PRR,CYT2024PRA}, or photonic synthetic dimensions~\cite{DLprl,YuanGA2022}. 

Thanks to the self-interference effects that profoundly modify their decay rates, frequency shifts, and coherence properties~\cite{fiveyears,LambAFK,SAW2014}, giant atoms exhibit a range of peculiar quantum phenomena, including decoherence-free (DF) interactions~\cite{NoriGA2018,OliverBraided,FCdeco,complexDFI,AFKchiral,AFKstructured,AFK2D,Leonforte2024}, chiral spontaneous emission~\cite{WXchiral2,DLprl,JoshiPRX2023,CYT2023PRR,DoublonWX}, non-Markovian retardation effects~\cite{GLZ2017,nonexp,DLprrTDcoupling,Guo2024NJP,ShenGA2024}, and unconventional bound states~\cite{WXchiral1,GuoOBS,Lim2023PRA}. 
Moreover, by engineering the band structure and Hermiticity of the environment, giant atoms can exhibit even more intriguing properties stemming from the interplay between self-interference effects and unconventional dispersion relations~\cite{Vega2021PRA,LeiQST,DLHN2023}. 
Despite these advances, the potential of giant-atom systems for manipulating and routing \textit{multi-qubit} states in a compact, scalable, and reliable manner remains largely unexplored.

\begin{figure}[hbt]
\centering
\includegraphics[width=\linewidth]{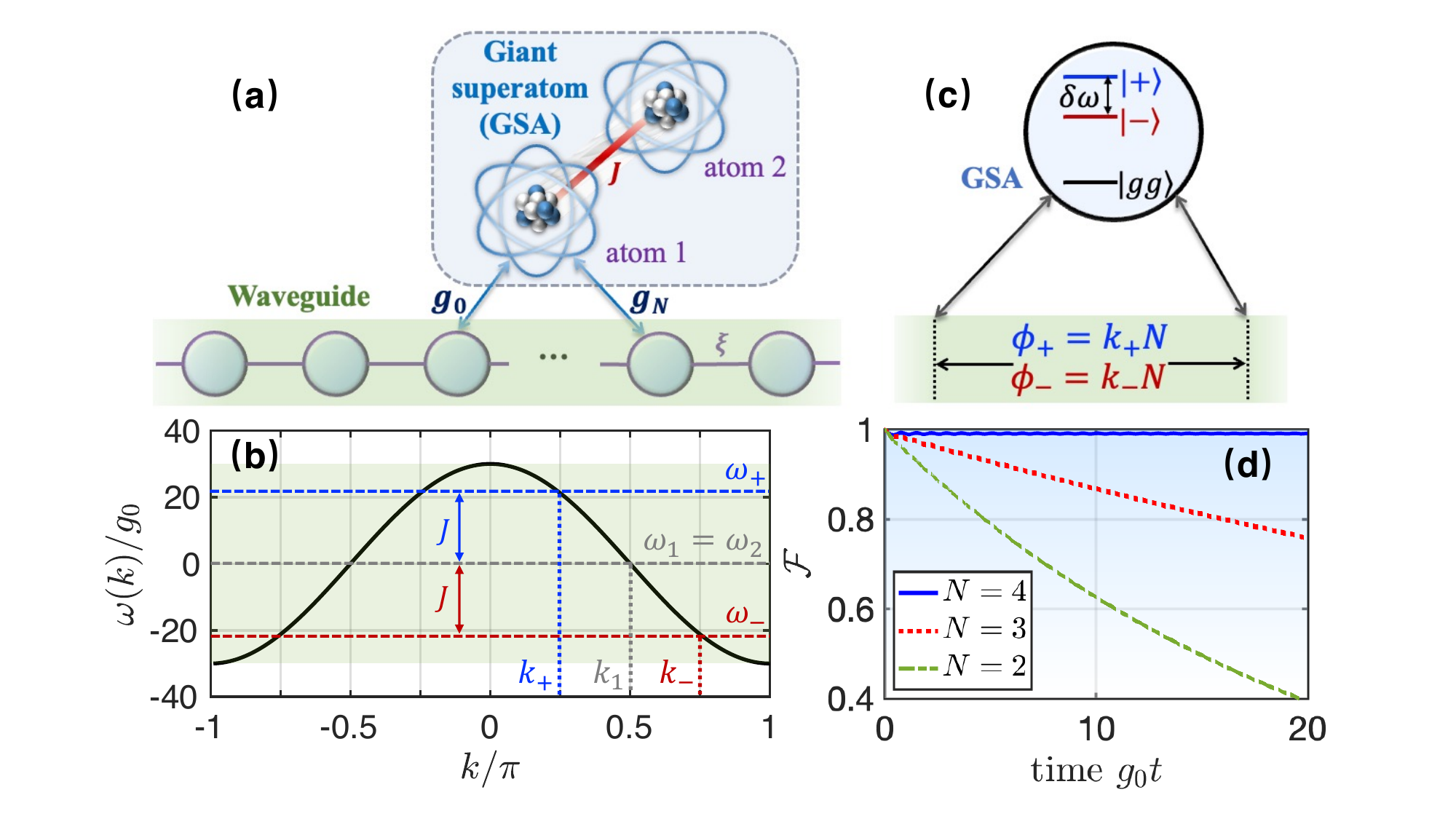}
\caption{(a) Schematic of a bipartite GSA, formed by a giant atom (atom 1) directly coupled to an additional atom (atom 2) via an interaction $J$. (b) Dispersion relation of the 1D tight-binding chain (modeling the waveguide). Colored lines indicate relevant frequencies and the corresponding wave vectors. (c) Dressed energy levels of the GSA and corresponding phase accumulations. (d) Time evolution of the fidelity $\mathcal{F}$ of $|\psi(t)\rangle$ with respect to $|\psi(0)\rangle$ for different values of $N$. Other parameters are $\omega_{1}=\omega_{2}=0$, $g_{N}/g_{0}\equiv1$, $\xi/g_{0}=15$, $J=\sqrt{2}\xi$, and $|\psi(0)\rangle=\mleft(\sigma_{+}^{(1)}+\sigma_{+}^{(2)}\mright)|G\rangle/\sqrt{2}$.}\label{FigModel}
\end{figure}

In this Letter, we fill this gap by introducing the concept of \emph{giant superatoms} (GSAs), i.e., composite systems consisting of two or more entangled atoms, collectively coupled to a waveguide through only one of them, at two or more separate coupling points; see Fig.~\ref{FigModel}{(a)}. Such systems host complex internal collective effects that can greatly extend the potential of giant-atom physics, while maintaining simple atom-field coupling structures. Unlike prior works~\cite{EntangleGA2023PRL,EntangleGA2023PRA,EntangleGA2023QIP,EantangleGA2024CPB,EntangleGA2025AQT}, this architecture encodes multi-qubit entanglement in a single compact node and addresses higher-dimensional entangled states without adding extra giant-atom units.
We reveal the dressed dynamics of GSAs and show how they lead to \emph{state-selective} interference effects. These interference effects enable a series of intriguing quantum-optical phenomena in both braided and separate GSA structures.
In braided structures, GSAs support DF transfer and swapping of internal dressed states. This mechanism can be further generalized to create various highly entangled states. For separate GSAs, we demonstrate state-selective chiral emission, which facilitates high-fidelity entanglement distribution over long distances and controlled generation of multipartite entangled states, such as $W$-class states. These results open new avenues for designing hybrid quantum systems, where nonlocal light-matter interactions and internal degrees of freedom are harnessed as key resources for quantum technologies.

\emph{Basic model.---}As illustrated in Fig.~\figpanel{FigModel}{a}, we begin by considering a bipartite GSA, which consists of a pair of two-level atoms and is coupled to a waveguide [represented here as a one-dimensional (1D) tight-binding chain] through atom 1 at two different points. The two atoms interact directly, enabling excitation exchange at a rate $J$. The Hamiltonian of the whole system is given by $H_{\text{tot}}=H_\mathrm{GSA}+H_\mathrm{chain}+H_{\text{int}}$, with ($\hbar=1$ in this Letter) 
\begin{subequations}
\begin{eqnarray}
H_\mathrm{GSA} &=& \sum_{l=1, 2}\omega_{l}\sigma_{+}^{(l)}\sigma_{-}^{(l)}+J\mleft(\sigma_{-}^{(1)}\sigma_{+}^{(2)}+\text{H.c.}\mright), \label{Hmol}\\ 
H_\mathrm{chain} &=& \sum_{j}\xi\mleft(a_{j}a_{j+1}^{\dag}+\text{H.c.}\mright), \label{Hbath}\\
H_{\text{int}} &=& \mleft(g_{0}a_{0}+g_{N}a_{N}\mright)\sigma_{+}^{(1)}+\text{H.c.}. \label{Hint}
\end{eqnarray}
\end{subequations}
The first part, $H_\mathrm{GSA}$, is the Hamiltonian of the GSA, where the raising (lowering) operator of atom $l$ is denoted by $\sigma_{+}^{(l)}$ ($\sigma_{-}^{(l)}$). The second part, $H_\mathrm{chain}$, describes the 1D tight-binding chain, with $a_{j}$ ($a_{j}^{\dag}$) the annihilation (creation) operator of the $j$th lattice site and $\xi$ the nearest-neighbor hopping rate. The dispersion relation $\omega(k)=2\xi\cos(k)$ of the 1D chain is depicted in Fig.~\figpanel{FigModel}{b}. Here we have set the (uniform) on-site potential of the lattice as the energy reference, so that $\omega_{l}$ represents the detuning of atom $l$ from the lattice band center. The final term, $H_{\text{int}}$, gives the interaction between atom 1 and two lattice sites $a_{0}$ and $a_{N}$, with coupling coefficients $g_{0}$ and $g_{N}$, respectively. Although a discrete waveguide (i.e., a 1D chain) is considered here, we point out that the results in this Letter also hold for the case of continuous waveguides. 

A single two-level giant atom features self-interference effects that are closely related to its transition frequency and the coupling-point separation~\cite{fiveyears,LambAFK,WilsonPRA2021}. Indeed, when $J=0$ and in the Markovian limit (i.e., with negligible retardation effect), the effective decay rate of the giant atom takes the form $\gamma_{\text{eff}}\propto g_{0}^{2}+g_{N}^{2}\cos(\phi)$, where $\phi=k_{1}N$ is the phase accumulation of the field between the two coupling points, with $k_{1}$ the wave vector corresponding to $\omega_{1}$ [see Fig.~\figpanel{FigModel}{b}]. The giant atom can thus exhibit either enhanced or suppressed decay to the waveguide, depending on the value of $\phi$. In particular, when $g_{0}=g_{N}$ and $\phi=\pi$ (mod $2\pi$), the giant atom retains a finite population in the long-time limit, forming a bound state in the continuum that is localized between its two coupling points. 

\textit{Working principle of GSAs.---}In the GSA proposed above, the interatomic coupling $J$ hybridizes the two constituent atoms, see Eq.~\eqref{Hmol}, such that it exhibits state-selective self-interference effects that depend on the value of $J$. Even when focusing on the single-excitation subspace, the unique features of GSAs can be well revealed.
Specifically, as illustrated in Fig.~\figpanel{FigModel}{c}, the GSA has two dressed eigenstates, $|+\rangle = \cos(\theta)|eg\rangle + \sin(\theta)|ge\rangle$ and $|-\rangle = \sin(\theta)|eg\rangle - \cos(\theta)|ge\rangle$, when $J\neq0$, with the mixing angle $\theta$ given by $\tan(2\theta)=2J/(\omega_{1}-\omega_{2})$. 
To lighten notation, we define the atomic states as, e.g., $|eg\rangle=|e\rangle_{1}\otimes|g\rangle_{2}$, which represents that atom 1 is in the excited state while atom 2 is in the ground state. The states $|\pm\rangle$ form an entangled basis within the single-excitation subspace; they have the corresponding eigenvalues 
\begin{equation}
\omega_{\pm}=\frac{\omega_{1}+\omega_{2}}{2}\pm\sqrt{\frac{\mleft(\omega_{1}-\omega_{2}\mright)^{2}}{4}+J^{2}}. \label{omegapm}
\end{equation}
When $g_{0}=g_{N}$, their effective decay rates are given by~\cite{BlaisPRA2011,SettineriPRA2018}
\begin{equation}
\Gamma_{\text{eff},\pm}=2\pi \mathcal{D}(\omega_{\pm})g_{0}^{2}[1+\cos{(\phi_{\pm})}]|s_{\pm}|^{2},
\label{effdecaypm}
\end{equation}
where $\mathcal{D}(\omega)$ is the density of states of the waveguide at frequency $\omega$, $\phi_{\pm}=k_{\pm}N$ are the associated phase accumulations with $k_{\pm}$ the wave vectors corresponding to $\omega_{\pm}$ [see Fig.~\figpanel{FigModel}{b}], and $s_{\pm}=\langle gg|\sigma_{-}^{(1)}|\pm\rangle$ [see Supplementary Material~\cite{SuppMat} for details and for generalization of Eq.~(\ref{effdecaypm}) to multipartite GSAs]. Note that a GSA differs fundamentally from a pair of giant atoms: in the latter case, the effective interatomic interactions \textit{arise} from the self-interference effects, while in GSAs the direct interatomic interactions actively modify the dynamics. Our proposal also behaves differently from coupled small atoms sharing a common waveguide~\cite{Kannan2023GM,Almanakly2025}, as further discussed in~\cite{SuppMat}. 

Now we consider a typical case where $\omega_{1}=\omega_{2}$. In this case, $|\pm\rangle$ are the symmetric and anti-symmetric entangled states in the single-excitation subspace, i.e., $|\pm\rangle=(|eg\rangle\pm|ge\rangle)/\sqrt{2}$, with eigenvalues $\omega_{\pm}=\omega_{1}\pm J$. Equation~(\ref{effdecaypm}) shows that $|\pm\rangle$ can exhibit self-interference effects that are distinct from those of a single giant atom. 
This difference arises from the state-selective phase accumulations, i.e., $\phi_{\pm}=k_{\pm}N$. For instance, when $\omega_{1}=0$ and $g_{0}=g_{N}$, the phase accumulation $\phi$ for a single giant atom is $\phi=N\pi/2$~\cite{AFKstructured}, such that it exhibits DF dynamics ($\gamma_{\text{eff}}=0$) for $N=2$ and shows enhanced decay ($\gamma_{\text{eff}}\propto 2g_{0}^{2}$) for $N=4$. 
In contrast, for the GSA with $J=\sqrt{2}\xi$, the phases become $\phi_{+}=N\pi/4$ and $\phi_{-}=3N\pi/4$ [see Fig.~\figpanel{FigModel}{b}], so that both $|+\rangle$ and $|-\rangle$ are dark simultaneously ($\Gamma_{\text{eff},\pm}=0$) when $N=4$ and decay rapidly into the waveguide when $N=2$.

In Fig.~\figpanel{FigModel}{d}, we provide a proof-of-principle demonstration of the self-interference effects of these entangled states (see \cite{SuppMat} for the numerical and analytical methods). The GSA is initialized in the symmetric entangled state, i.e., $|\psi(0)\rangle=\mleft(\sigma_{+}^{(1)}+\sigma_{+}^{(2)}\mright)|G\rangle/\sqrt{2}$, where $|G\rangle$ denotes the global ground state of the whole setup (with all atoms in their ground states and the waveguide in vacuum) throughout this work. We then track the time evolution of the fidelity $\mathcal{F}(t)=|\langle\psi(t)|\psi(0)\rangle|$, where $|\psi(t)\rangle=\mleft[c_{1}(t)\sigma_{+}^{(1)}+c_{2}(t)\sigma_{+}^{(2)}\mright]|G\rangle$ is the state at time $t$ with superposition coefficients $c_{1,2}(t)$. As expected, the entangled state remains dark when $N=4$ and exhibits enhanced decay as $N$ reduces to $2$. As will be shown below, this effect enables a \emph{DF swapping} between the symmetric and anti-symmetric entangled states in braided GSA structures.

\begin{figure}[ptb]
\centering
\includegraphics[width=0.9\linewidth]{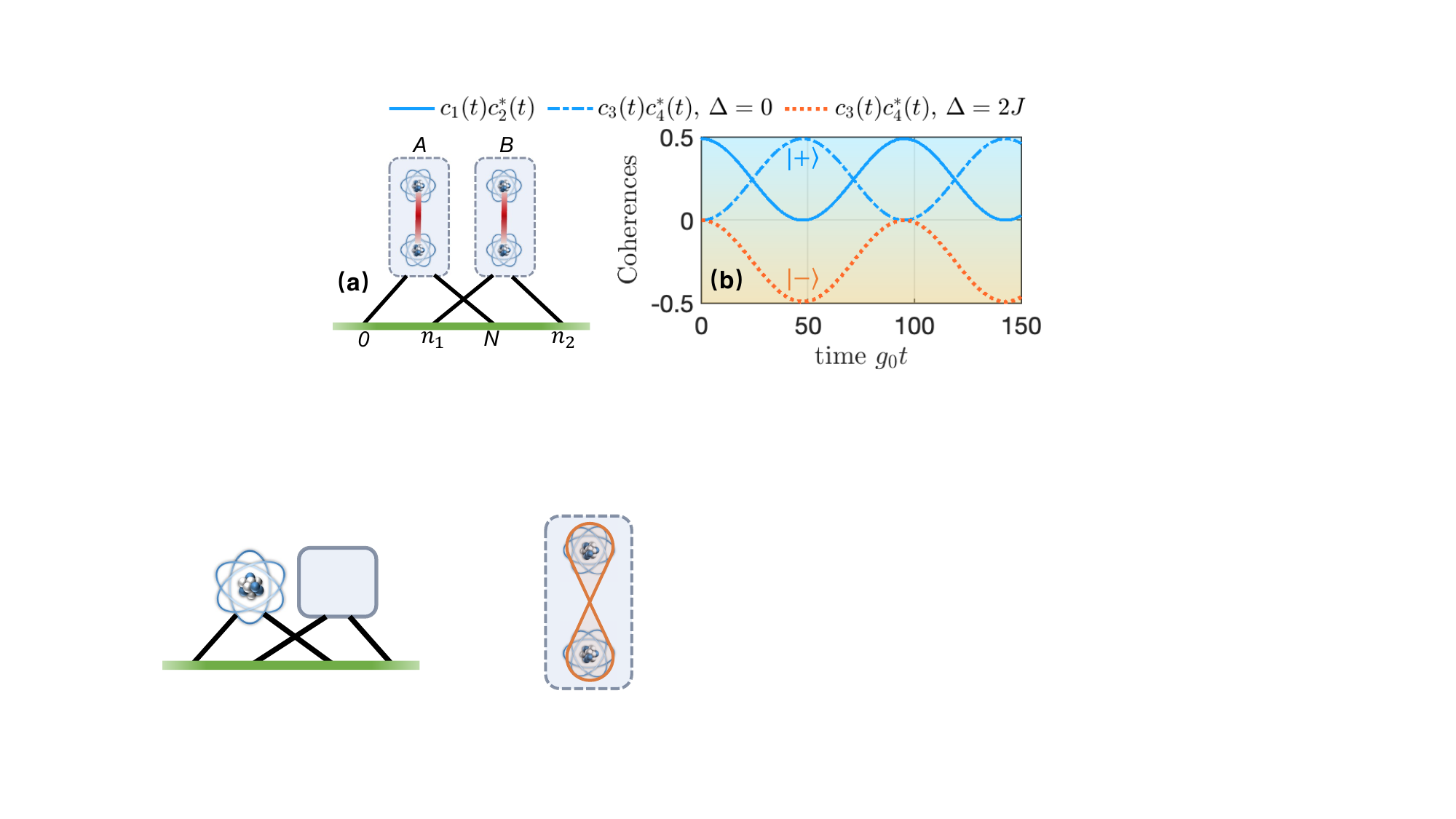}
\caption{(a) Schematic of the braided GSA structure. (b) Time evolution of the atomic coherences $c_{l}(t)c_{l'}^{*}(t)$ for two different values of $\Delta$, starting from an initial state $|\psi(0)\rangle=\mleft(\sigma_{+}^{(1)}+\sigma_{+}^{(2)}\mright)|G\rangle/\sqrt{2}$. 
Other parameters are $\omega_{1}=\omega_{2}=\omega_{3}-\Delta=\omega_{4}-\Delta=0$, $g_{0}=g_{n_{1}}=g_{N}=g_{n_{2}}$ with $\{n_{1}, N, n_{2}\}=\{1, 4, 5\}$, $\xi/g_{0}=15$, and $J=J'=\sqrt{2}\xi$.}\label{FigDFI}
\end{figure}

\emph{Braided structures.---}We now extend the above model to a braided coupling structure, where a second GSA (labeled ``GSA $B$'') is coupled to the waveguide with its two coupling points (with coupling coefficients $g_{n_{1}}$ and $g_{n_{2}}$) between and outside the two coupling points of the present one (labeled ``GSA $A$''), respectively, see Fig.~\figpanel{FigDFI}{a}. We therefore include two additional Hamiltonian terms $H_\mathrm{GSA}'=\sum_{l=3, 4}\omega_{l}\sigma_{+}^{(l)}\sigma_{-}^{(l)}+J'(\sigma_{-}^{(3)}\sigma_{+}^{(4)}+\text{H.c.})$ and $H_{\text{int}}'=(g_{n_{1}}a_{n_{1}}+g_{n_{2}}a_{n_{2}})\sigma_{+}^{(3)}+\text{H.c.}$ in $H_{\text{tot}}$ to describe this extended model, where $J'$ is the interatomic coupling strength of GSA $B$ and the coupling positions satisfy $0<n_{1}<N<n_{2}$. Accordingly, the state of the whole setup at time $t$ can be written as $|\psi(t)\rangle=\sum_{l=1}^{4}c_{l}(t)\sigma_{+}^{(l)}|G\rangle$. 

We now show how this braided structure can be exploited for DF entanglement \emph{transfer} and \emph{swapping}. In Fig.~\figpanel{FigDFI}{b}, we also plot the time evolution of the coherences $c_{l}(t)c_{l'}^{*}(t)$ between pairs of atoms, with the requirement that the $|\pm\rangle$ states of both GSAs are dark [cf.~the cyan solid line in Fig.~\figpanel{FigModel}{d}]. When the two GSAs have the same frequency, they exhibit a DF interaction: the initial symmetric entangled state of GSA $A$ can be transferred to GSA $B$ with nearly negligible decoherence (extremely weak decoherence occurs due to non-Markovian retardation effects~\cite{LonghiGA,complexDFI,AFKstructured}). 
This DF mechanism can be analytically understood using the resolvent formalism, as detailed in~\cite{SuppMat}. Interestingly, when GSA $B$ is detuned from GSA $A$ by $\Delta=\omega_{3,4}-\omega_{1,2}=2J$ with $J=\sqrt{2}\xi$, the initial symmetric entangled state of GSA $A$ is converted into the anti-symmetric entangled state of GSA $B$, leading to a DF swapping of entangled states. This result can be understood from the fact that, when $\Delta=2J$, the ``$+$'' state of GSA $A$ becomes resonant with the ``$-$'' state of GSA $B$, both of which satisfy the DF condition.  

\begin{figure}[ptb]
\centering
\includegraphics[width=\linewidth]{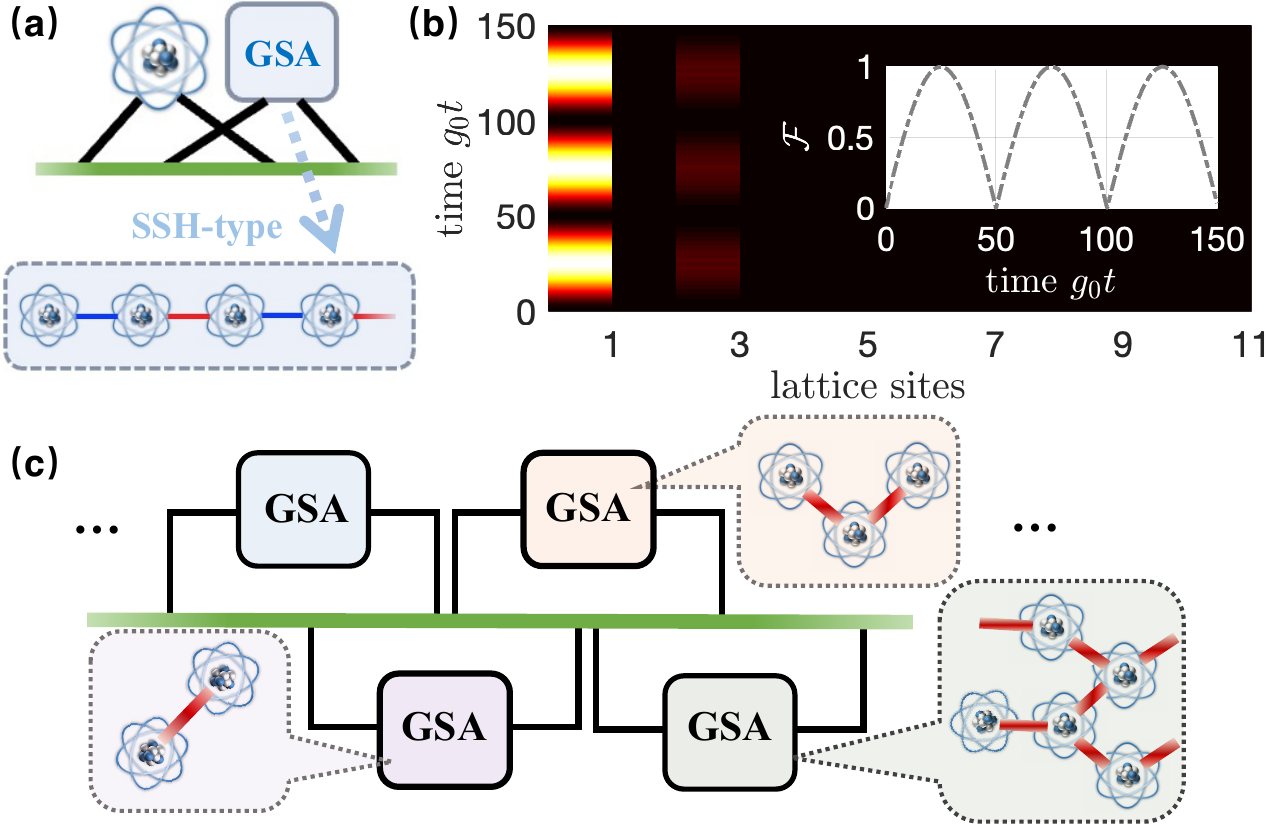}
\caption{(a) Schematic of an extended braided structure, where a giant atom is braided with an SSH-type GSA. (b) Time evolution of the excitation probability distribution in the SSH-type GSA with $M=6$ and $\omega_{1}=0$. Inset: time evolution of the fidelity $\mathcal{F}$ of the transferred state with respect to the topological left edge state of the GSA, assuming the giant atom is initially excited. Other parameters are $g_{0}=g_{n_{1}}=g_{N}=g_{n_{2}}$ with $\{n_{1}, N, n_{2}\}=\{1, 2, 3\}$, $\xi/g_{0}=15$, $J_{1}/g_{0}=0.5$, and $J_{2}/g_{0}=1.5$. (c) Schematic of a ``structured entanglement lattice'' formed by a chain of braided GSAs. Each lattice site encodes an entangled state of the constituent atoms, rather than a single-atom excitation.}\label{FigSSH}
\end{figure}

The above mechanism can be further extended to generate more complex superposition states by incorporating GSAs composed of more atoms. In fact, one can realize DF swapping between arbitrary eigenstates of two braided giant subsystems, provided that the target states have the same energy while remaining distinguishable from all other eigenstates of the subsystems. As shown in Fig.~\figpanel{FigSSH}{a}, for example, we replace GSA $A$ in Fig.~\figpanel{FigDFI}{a} with a single giant atom (i.e., setting $J=0$), and replace GSA $B$ with a Su--Schrieffer--Heeger (SSH) chain~\cite{SSH1979,SSH1980} of two-level atoms. 
In this case, the total Hamiltonian reads $\tilde{H}_{\text{tot}}=\omega_{1}\sigma_{+}^{(1)}\sigma_{-}^{(1)}+H_\mathrm{chain}+H_{\text{SSH}}+H_{\text{int}}+\tilde{H}_{\text{int}}$, where $H_\mathrm{chain}$ and $H_{\text{int}}$ are given in Eqs.~(\ref{Hbath}) and (\ref{Hint}), respectively, and 
\begin{subequations}
\begin{eqnarray}
H_{\text{SSH}} & = &\sum_{l=1}^{M}J_{1}\sigma_{+}^{P,l}\sigma_{-}^{Q,l}+\sum_{l=1}^{M-1}J_{2}\sigma_{+}^{P,l+1}\sigma_{-}^{Q,l}+\text{H.c.}, \label{Hssh}\\
\tilde{H}_{\text{int}} & = & \mleft(g_{n_{1}}a_{n_{1}}+g_{n_{2}}a_{n_{2}}\mright)\sigma_{+}^{P,1}+\text{H.c.}. \label{Hinttilde}
\end{eqnarray}
\end{subequations}
Here $\sigma_{+}^{P,l}=(\sigma_{-}^{P,l})^{\dag}$ and $\sigma_{+}^{Q,l}=(\sigma_{-}^{Q,l})^{\dag}$ are, respectively, the raising operators of the two sublattice sites (labeled $P$ and $Q$) in the $l$th unit cell of the SSH model. $J_{1}$ and $J_{2}$ are the associated intracell and intercell coupling strengths, respectively. 
A finite SSH model, as described by Eq.~(\ref{Hssh}), is known to host two topological edge states in the topologically nontrivial phase (i.e., when $J_{1}<J_{2}$)~\cite{TopoInsulators}. These edge states are zero modes with eigenvalues pinned to the center of the middle energy gap. Despite being degenerate, the two edge states are mutually orthogonal since they occupy different sublattices. Consequently, we anticipate DF swapping between the giant-atom excitation and a given edge state under suitable conditions. Figure~\figpanel{FigSSH}{b} shows the high-fidelity generation of a topological edge state that is strongly localized on the left end of the SSH-type GSA. The process begins with the giant atom initially prepared in the excited state. This selective swapping is achieved by coupling the leftmost (i.e., first) atom of the GSA to the waveguide, such that only the left edge state interacts with the giant-atom excitation, while its effective coupling strength to the waveguide is simultaneously maximized. This result demonstrates the versatility and scalability of our DF manipulation scheme.

Furthermore, as schematically illustrated in Fig.~\figpanel{FigSSH}{c}, a chain of braided GSAs can be employed to realize a \emph{structured entanglement lattice}, in which each site encodes an entangled state of multiple atoms, rather than a simple single-atom excitation. This architecture marks a significant departure from conventional tight-binding lattices: its composite sites possess internal correlations and are coherently coupled via protected waveguide-mediated interactions. The DF mechanism preserves coherence during this process, thereby facilitating the transfer and redistribution of these entangled states across the lattice. Further elaboration on this proposal is provided in Appendix~A.

\begin{figure}[ptb]
\centering
\includegraphics[width=8.5 cm]{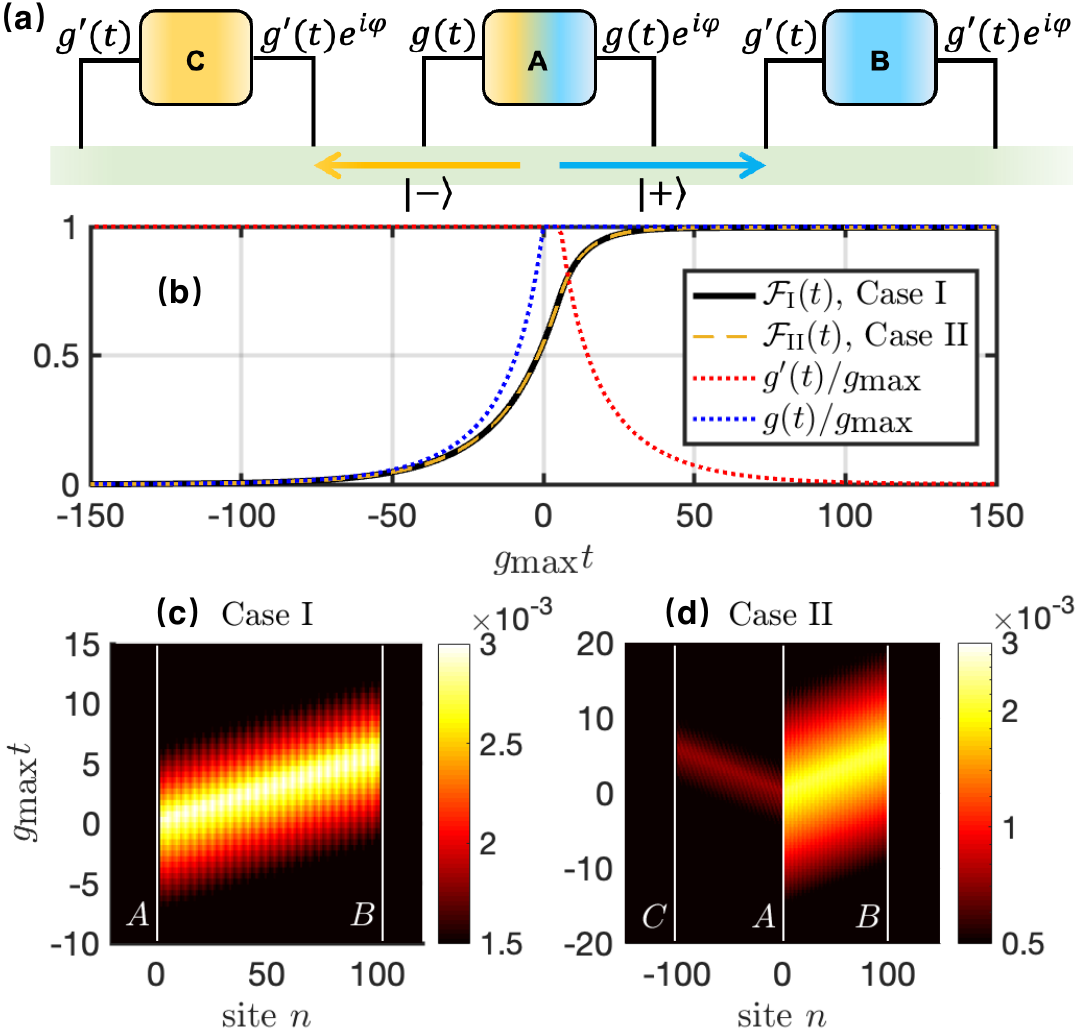}
\caption{(a) Schematic of the separate GSA structure. High-efficiency chiral state transfer between remote GSAs is allowed by engineering the coupling phase difference $\varphi$. (b) The time-dependent coupling coefficients $g(t)$ and $g'(t)$ together with the time evolution of the fidelities $\mathcal{F}_{\mathrm{I}}$ and $\mathcal{F}_{\mathrm{II}}$. (c) and (d) Time evolution of the field intensity distribution $|a_{n}(t)|^{2}$ for cases I and II (see text for details). Other parameters are $\omega_{j=1,2,3,4}\equiv0$, $N=2$, $n_{1}=100$, $n_{2}=102$, $n_{3}=-102$, $n_{4}=-100$, $\xi/g_{\text{max}}=12.5$, $J=J'=\sqrt{2}\xi$, $\varphi=\pi/2$, $g_{\text{max}}\tilde{\tau}=5.657$, and $\beta/g_{\text{max}}=0.045$.}\label{FigChiral}
\end{figure}

\emph{Separate structures.---}Up to now, we have focused on braided coupling structures, which enable DF dynamical processes, yet require the giant (super)atoms to be close to each other. A long-standing challenge in quantum technologies has been to achieve deterministic state transfer between \textit{remote} nodes~\cite{Kimble2008}. In giant-atom systems, chiral spontaneous emission, i.e., directionally biased photon emission~\cite{Lodahl2017nature}, can be achieved by carefully engineering the local coupling phases at multiple coupling points~\cite{WXchiral2,DLprl,JoshiPRX2023}. This makes such systems promising building blocks for high-fidelity remote state transfer.  

For a bipartite GSA, one can hence naturally expect distinct chiralities for the states $|\pm\rangle$. As illustrated in Fig.~\figpanel{FigChiral}{a}, this is achieved by tuning the phase difference $\varphi$ between the two coupling coefficients of the GSA to break the time-reversal symmetry of the effective atom-field interaction. In particular, $|+\rangle$ and $|-\rangle$ exhibit opposite chiralities when $\varphi-\phi_{+}$ ($\varphi-\phi_{-}$) is an even (odd) multiple of $\pi$ while $\varphi+\phi_{+}$ ($\varphi+\phi_{-}$) is an odd (even) multiple of $\pi$~\cite{SuppMat}. As a result, excitations in $|\pm\rangle$ can be emitted in opposite directions and subsequently be reabsorbed by another quantum node positioned on the corresponding side of this GSA. 

We first explore the possibility of \textit{state transfer} (referred to as ``case I'') where two GSAs, $A$ and $B$, are spatially \emph{separate}, as shown in the right part of  Fig.~\figpanel{FigChiral}{a} (disregarding GSA $C$ for the moment), with $\omega_{j}=0$ ($j=1,2,3,4$), $J=J'=\sqrt{2}\xi$, and $0<N<n_{1}<n_{2}$. We also assume an identical phase difference $\varphi=\pi/2$ for both GSAs, which can be readily achieved using SQUID couplers and appropriate modulations of the magnetic fluxes~\cite{SuppMat}. 
To achieve high-efficiency state transfer, the atom-field coupling coefficients should be carefully modulated to eliminate the back reflection of the field from the receiving GSA~\cite{WXchiral2,Cirac1997Transfer,KorotkovPRB2011}. Without loss of generality, we assume identical time-dependent coupling coefficients $g(t)$ [$g'(t)$] for the two coupling points of GSA $A$ (GSA $B$), which satisfy
\begin{equation}
g(t) = g'(\tilde{\tau}-t) = 
\begin{cases} 
g_{\text{max}}\frac{e^{\beta t}}{2 - e^{\beta t}}, & t < 0, \\ 
g_{\text{max}}, & t \geq 0.
\end{cases}
\label{chiralgt}
\end{equation}
Here, $\tilde{\tau}$ (see~\cite{SuppMat} for the explicit expression) is the propagation time of the field between the GSAs, while $g_{\text{max}}$ and $\beta$ determine the maximum and the slope of the coupling modulation, respectively. Figure~\figpanel{FigChiral}{b} shows the evolution of $g(t)$ and $g'(t)$, which obey a time-reversal symmetry with respect to $t=\tilde{\tau}/2$. We also compute the fidelity $\mathcal{F}_{I}(t)$ of the evolved state, which is initialized as $|\psi(0)\rangle_{\mathrm{I}}=\mleft(\sigma_{+}^{(1)}+\sigma_{+}^{(2)}\mright)|G\rangle/\sqrt{2}$, with respect to the target state $\mleft(\sigma_{+}^{(3)}+\sigma_{+}^{(4)}\mright)|G\rangle/\sqrt{2}$. It shows that the fidelity approaches unity ($>\SI{99}{\percent}$) with appropriate values of $\beta$~\cite{SuppMat}. In Fig.~\figpanel{FigChiral}{c}, we plot the evolution of the field intensity distribution $|a_{n}(t)|^{2}$, which confirms the nearly perfect directional state transfer between the two GSAs, as expected. Note that high-fidelity chiral transfer is also achievable for general specified superposition states---beyond the Bell states considered above; see \cite{SuppMat} for further details. 

Importantly, \textit{multipartite entanglement generation} can be achieved with remote GSAs, if one considers a more general case (referred to as ``case II'') where GSA $A$ is initially prepared in a superposition $c_{+}|+\rangle+c_{-}|-\rangle$, i.e.,
\begin{equation}
|\psi(0)\rangle_{\mathrm{II}} 
= \mleft(c_{+}\frac{\sigma_{+}^{(1)}+\sigma_{+}^{(2)}}{\sqrt{2}}+c_{-}\frac{\sigma_{+}^{(1)}-\sigma_{+}^{(2)}}{\sqrt{2}}\mright)|G\rangle
\label{entangle0}
\end{equation}
with $|c_{+}|^{2}+|c_{-}|^{2}=1$. In this case, the $|+\rangle$ and $|-\rangle$ components can be emitted in opposite directions, such that the original entanglement in GSA $A$ is transferred to photonic modes propagating in different directions, thereby creating spatially separated entangled states~\cite{SimonePRL2022}. Accordingly, we introduce a third GSA (labeled $C$), which is identical to GSA $B$---including the interatomic coupling $J'$ and the time-dependent coupling coefficient $g'(t)$---but positioned on the opposite side of GSA $A$ (with its two coupling points located at $a_{n_{3}}$ and $a_{n_{4}}$); see \cite{SuppMat} for details. Following the modulation scheme in Eq.~(\ref{chiralgt}), the $|+\rangle$ and $|-\rangle$ components of $|\psi(0)\rangle_{\mathrm{II}}$ can be deterministically transferred to GSAs $B$ and $C$, respectively. As a result, GSAs $B$ and $C$ finally form a $W$-class entangled state~\cite{CiracGHZ2000,WstateArxiv2008,WstatePRA2020} 
\begin{equation}
|\psi_{T}\rangle
= \frac{c_{+}}{\sqrt{2}} \mleft(|eggg\rangle + |gegg\rangle\mright) + \frac{c_{-}}{\sqrt{2}} \mleft(|ggeg\rangle - |ggge\rangle\mright),
\label{entanglef}
\end{equation}
where the lattice field and GSA $A$ have been traced out. Here, we use simplified notation, e.g., $|eggg\rangle=|eg\rangle_{B}\otimes|gg\rangle_{C}$, to represent the states of GSAs $B$ and $C$. Starting from the initial state in Eq.~(\ref{entangle0}), with $c_{+}=\sqrt{3}/2$ and $c_{-}=1/2$, the evolution of the corresponding fidelity $\mathcal{F}_{\mathrm{II}}$ and the field intensity distribution are depicted in Figs.~\figpanel{FigChiral}{b} and \figpanel{FigChiral}{d}, respectively. As expected, a high-fidelity $W$-state is generated between GSAs $B$ and $C$, even at a large separation distance. In Appendix~B, we further discuss how to extend this protocol to involve more complex GSA configurations.

\emph{Conclusion.---}We have introduced the concept of GSAs as a new quantum optical paradigm. Specifically, we demonstrated a series of quantum processes in single, braided, and separate GSA configurations, including dressed self-interference effects, DF state transfer and swapping, and state-dependent chiral spontaneous emission. In particular, the latter makes it possible to achieve selective, directional transfer of quantum information and remote generation of $W$-class entangled states. These findings highlight the potential of GSAs in quantum communication and quantum networks.

Our work opens several promising directions for future research. For instance, engineering GSAs in topological structured baths~\cite{Vega2021PRA,ChengSSH2022,Vega2023PRR,Leonforte2024} could enable robust entanglement generation and protection against decoherence, crucial for scalable quantum networks. Moreover, the interplay between GSAs and non-Hermitian photonic lattices~\cite{DLHN2023,DuHN2025} offers an avenue to engineer exotic decay dynamics and non-reciprocal quantum transport. By leveraging structured light-matter interactions, self-interference effects, and internal degrees of freedom, GSAs provide a versatile platform for programmable quantum information processing.

\acknowledgments{We thank F.~Nori, A.~C. Santos, P.~Rabl, M.~Horodecki, and A.~Soro for helpful discussions.  X.W.~is supported by the National Natural Science Foundation of China (NSFC) (Grant No.~12174303).
A.F.K.~acknowledges support from the Swedish Foundation for Strategic Research (grant numbers FFL21-0279 and FUS21-0063), the Horizon Europe programme HORIZON-CL4-2022-QUANTUM-01-SGA via the project 101113946 OpenSuperQPlus100, and from the Knut and Alice Wallenberg Foundation through the Wallenberg Centre for Quantum Technology (WACQT). J.S. and L.D.~acknowledge financial support by the Knut och Alice Wallenberg stiftelse through project grant no. 2022.0090, as well as through an individual Wallenberg Academy fellowship grant.
}

\emph{Data availability.---}The data that support the findings of this article are not publicly available. The data are available from the authors upon reasonable request.


\bibliography{sample}


\section*{End Matter}\label{EndMatter}
\renewcommand{\thesection}{EM\arabic{section}} 

\begin{figure}[pth]
\centering
\includegraphics[width=7 cm]{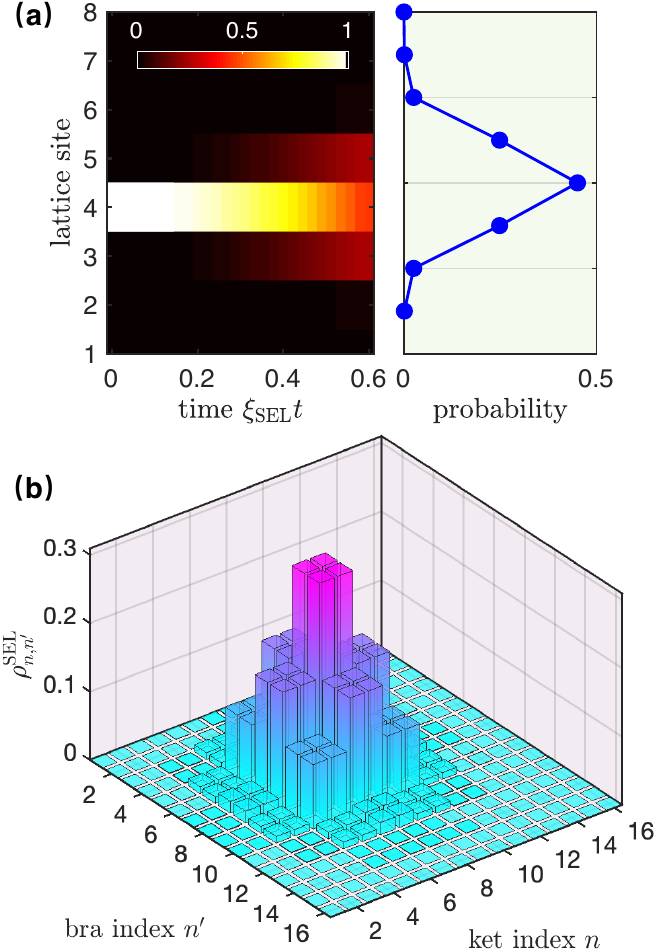}
\caption{(a) Time evolution of the probability distribution (left) and distribution profile at the final time (right) of a structured entanglement lattice, which is formed by a chain of braided bipartite GSAs. The process begins with the fourth GSA prepared in its $|+\rangle$ dressed state. (b) State tomography (density matrix elements $\rho_{n,n'}^{\mathrm{SEL}}$) of the $16$ physical atoms forming the structured entanglement lattice, which indicates a nonlocal multipartite entanglement.   
}\label{SMsel}
\end{figure}

\emph{Appendix~A: Dynamical entanglement spreading in a structured entanglement lattice.---}In Fig.~\figpanel{FigSSH}{c}, we introduced the concept of a ``structured entanglement lattice'' formed by a chain of braided GSAs. Here, we illustrate the dynamical spreading of entanglement in such a lattice, showing how an initially \emph{local} entangled state can evolve into a \emph{nonlocal} entangled state involving an increasing number of physical qubits. As a simple example, we consider a uniform 1D chain of braided GSAs, each composed of two atoms as in Fig.~\ref{FigModel}. We initialize one of these GSAs (e.g., the fourth one) in its $|+\rangle$ state, and let this excitation propagate along the GSA chain. The dynamics are governed by the effective tight-binding Hamiltonian
\begin{equation}
H_{\mathrm{SEL}}=\xi_{\mathrm{SEL}}\sum_{m}|+,m\rangle\langle+,m+1|+\mathrm{H.c.},
\end{equation}
where $|+,m\rangle$ denotes the ``+'' dressed state of the $m$th GSA, and the effective hopping rate is given by $\xi_{\mathrm{SEL}}\approx\Sigma_{\mathrm{eff},+}(0+i0^{+})=g_{0}^{2}/2\xi$, derived from Eq.~(S40) in \cite{SuppMat}.

Figure~\figpanel{SMsel}{a} shows the dynamics of the probability distribution across eight GSAs, along with the distribution profile at the final time. Since each site in this lattice model represents a bipartite entangled state of two physical atoms, the overall state at time $t$ is actually a coherent superposition of these local entangled states:
\begin{equation}
|\psi_{\mathrm{SEL}}(t)\rangle=\sum_{m}\mathcal{C}_{m}(t)|+,m\rangle,
\label{stateSEL}
\end{equation}
where $\mathcal{C}_{m}(t)$ is the probability amplitude of exciting the $m$th $|+\rangle$ state at time $t$. To visualize the entanglement distribution among the physical atoms, we rewrite the state in the original atomic basis as $|\psi_{\mathrm{SEL}}'(t)\rangle=\sum_{n}\mathcal{C}_{n}'(t)\sigma_{+}^{(n)}|G\rangle$, where $\mathcal{C}_{n}'(t)$ represents the excitation amplitude of the $n$th physical atom (two atoms per lattice site). For the case in Fig.~\figpanel{SMsel}{a}, these amplitudes satisfy $\mathcal{C}_{2m-1}'(t)=\mathcal{C}_{2m}'(t)=\mathcal{C}_{m}(t)/\sqrt{2}$. Figure~\figpanel{SMsel}{b} shows the elements of the density matrix $\rho^{\mathrm{SEL}}=|\psi_{\mathrm{SEL}}'\rangle\langle\psi_{\mathrm{SEL}}'|$, which captures the single-excitation sector of the full multipartite state, evaluated at the final time. The central peak and nonzero off-diagonal coherences clearly indicate that the initial localized bipartite entanglement has spread across the lattice, forming a more complex multipartite structure that correlates a larger number of atoms. Inspired by this result, one may also expect the periodic revival of such nonlocal multipartite entangled states by introducing a linear gradient in the on-site potentials, e.g., $F\sum_{m}m|+,m\rangle\langle+,m|$, which induces Bloch oscillations of the excitation~\cite{Bloch1999prl}.


\begin{figure}[pth]
\centering
\includegraphics[width=8.5 cm]{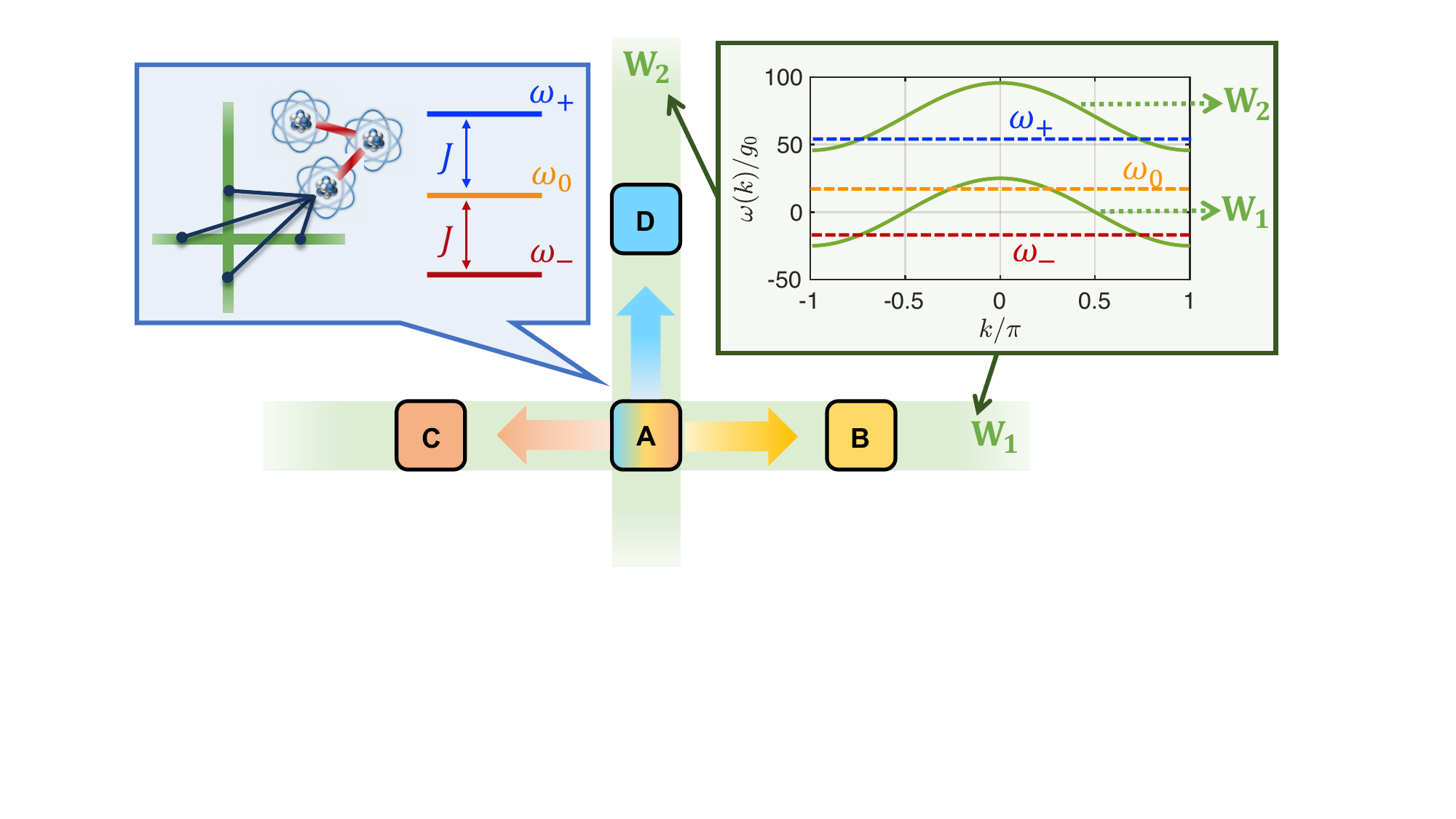}
\caption{Schematic of the extended entanglement generation scheme, involving more GSAs and routing channels. In this scheme, the emitting GSA ($A$) is composed of three atoms in a chain and thus exhibits three dressed states (with eigenfrequencies $\omega_{0}$ and $\omega_{\pm}$, respectively). GSA $A$ is nonlocally coupled to both waveguides simultaneously, via two separate coupling points each. The two lower eigenfrequencies, $\omega_{-}$ and $\omega_{0}$, lie within the energy band of waveguide $\text{W}_{1}$, while the upper one, $\omega_{+}$, lies within the band of waveguide $\text{W}_{2}$. The system is initialized with GSA $A$ in a superposition of the three dressed states. By engineering the phase differences among the coupling coefficients of GSA $A$ and applying appropriate coupling modulation schemes, each component of the initial superposition state can be selectively transferred to GSAs $B$, $C$, and $D$, respectively. 
}\label{SMextension}
\end{figure}

\emph{Appendix~B: Multi-channel state transfer protocols.---}The chiral state-transfer protocol in Fig.~\ref{FigChiral} can be naturally extended to support more complex configurations of GSAs and more routing channels. For instance, consider the extended setup shown in Fig.~\ref{SMextension}, where the emitting GSA (labeled $A$) is formed by a chain of three identical atoms with an interatomic coupling rate $J$. Such a GSA exhibits three distinct dressed states, 
\[
|\psi_{\pm}\rangle = \frac{1}{2}
\begin{pmatrix}
1 \\
\pm \sqrt{2} \\
1
\end{pmatrix} \quad \mathrm{and} \quad
|\psi_{0}\rangle = \frac{1}{\sqrt{2}}
\begin{pmatrix}
-1 \\
0 \\
1
\end{pmatrix},
\] 
with eigenfrequencies $\omega_{\pm}=\omega_{0} \pm \sqrt{2}J$ and $\omega_0$, respectively. GSA $A$ is nonlocally coupled to two waveguides, $\text{W}_1$ and $\text{W}_2$, through two spatially separated coupling points each (with identical coupling coefficient $g_{0}$). The spectral structure is engineered such that the lower two eigenfrequencies, $\omega_{-}$ and $\omega_{0}$, fall within the energy band of $\text{W}_1$, while the upper eigenfrequency $\omega_{+}$ lies within the energy band of $\text{W}_2$. The spectrum in Fig.~\ref{SMextension} (top right) shows an example with $\xi=12.5g_{0}$, $\omega_{0}=\sqrt{2}\xi$, and $J=2\xi$ (all symbols, except those specified in this paragraph, are identical to those in the main text). Here we take the band center of $\text{W}_{1}$ as the energy reference and assume a higher band center $4\sqrt{2}\xi$ for $\text{W}_{2}$.

By initializing GSA $A$ in a coherent superposition of its three dressed states, i.e., 
\begin{equation}
|\psi(0)\rangle_{A} = \sum_{\nu=\pm,0} c_{\nu}|\psi_{\nu}\rangle,
\label{initialPM0}
\end{equation}
and carefully designing the phase differences among its coupling coefficients, each component of the initial state can be routed into a distinct propagation direction. Consequently, the emitted wave packets are directed toward three spatially separated receiving GSAs (labeled $B$, $C$, and $D$), which are strategically positioned to selectively absorb a specific component, through appropriate coupling phase engineering and coupling amplitude modulation (e.g., similar to that used in Fig.~\ref{FigChiral}). Assuming that GSAs $B$, $C$, $D$ are identical in configuration to $A$, one can finally obtain a multipartite entangled state of the form
\begin{eqnarray}
|\psi_{T}\rangle_{BCD} &=& c_{0}|\psi_{0}\rangle_{B}\otimes|ggg\rangle_{C}\otimes|ggg\rangle_{D} \nonumber\\
&& + c_{-}|ggg\rangle_{B}\otimes|\psi_{-}\rangle_{C}\otimes|ggg\rangle_{D} \nonumber\\
&& + c_{+}|ggg\rangle_{B}\otimes|ggg\rangle\otimes|\psi_{+}\rangle_{D},
\label{finalPM0}
\end{eqnarray}
where the subscripts $B$, $C$, and $D$ indicate the states of the corresponding GSAs.



\end{document}